**Titre + Résumé**

**Stockage de charge dans les carbones nanoporeux: L'origine moléculaire de la supercapacité**

Très performants en puissance, les supercondensateurs sont par exemple utilisés pour récupérer l'énergie de freinage dans certaines voitures ou tramways. Ils fonctionnent grâce à deux électrodes en carbone plongées dans une solution ionique ou un liquide ionique pur. C'est l'adsorption d'ions à la surface des électrodes qui permet de stocker l'électricité, mais le mécanisme microscopique à l'origine des performances exceptionnelles des carbones dérivés de carbures (CDC) pour le stockage de la charge restait à établir. Par simulation moléculaire d'électrodes de structure réaliste et maintenues à potentiel constant, nous étudions les effets du confinement et de la solvatation sur le mécanisme de charge. Nous précisons également la dynamique du processus de charge et faisons le lien avec les modèles utilisés par les électro-chimistes.

**Mots-clés**

Supercondensateur, Simulation moléculaire, Adsorption, Echange ionique, Electrode, Carbone nanoporeux

**Title + Abstract**

**Charge storage in nanoporous carbons: The molecular origin of supercapacitance**

Supercapacitors are electric devices able to deliver a large power, enabling their use e.g. for the recovery of breaking energy in cars. This is achieved by using two carbon electrodes and an electrolyte solution or a pure ionic liquid (Room Temperature Ionic Liquid, RTIL). Energy is stored by the adsorption of ions at the surface of the electrodes, but the microscopic mechanism underlying the exceptional performance of Carbide Derived Carbon (CDC) electrodes remained unknown. Using molecular simulation with realistic electrode structures and under constant voltage conditions, we investigate the effect of confinement and solvation on the microscopic charging mechanism. We further analyse the dynamics of the charging process and make the link with equivalent circuit models used by electrochemists.

**Keywords**

**Auteurs**

Céline Merlet[1,2], Clarisse Péan[1,2], M. Salanne[1,2] et Benjamin Rotenberg[1,2]

1 Sorbonne Universités, UPMC et CNRS, Laboratoire PHENIX, 4 place Jussieu, 75005 Paris

2 RS2E (Réseau sur le Stockage Electrochimique de l'Energie)

Benjamin Rotenberg est chargé de recherche CNRS et Mathieu Salanne est maître de conférences au laboratoire Physicochimie des Electrolytes et Nanosystèmes Interfaciaux (PHENIX, UMR 8234) à l'UPMC (Université Pierre et Marie Curie, Paris). Céline Merlet et Clarisse Péan sont actuellement en post-doctorat, respectivement à l'Université de Cambridge (Royaume-Uni) et à Chimie ParisTech (Paris).

## 1) Introduction

Les supercondensateurs sont des dispositifs de stockage de l'électricité complémentaires aux batteries, qui permettent de stocker une grande quantité d'énergie mais ne délivrent qu'une faible puissance, et aux condensateurs conventionnels, qui ont des performances opposées (voir Figure 1) [1,2]. Les supercondensateurs peuvent ainsi se charger ou délivrer leur énergie en quelques secondes ou dizaines de secondes. Cette caractéristique des supercondensateurs explique leur utilisation dans un nombre croissant d'applications nécessitant une grande puissance, telle l'ouverture d'urgence des portes de l'avion Airbus A380, ou dans les dispositifs Stop&Start des véhicules. Ils permettent également de récupérer une partie de l'énergie cinétique perdue lors du freinage des véhicules comme les tramways ou de l'énergie potentielle lors de la descente des grues portuaires, avant sa réutilisation lors du redémarrage ou de la remontée.

Alors que les supercondensateurs sont en général utilisés de façon complémentaire à une autre source d'énergie, on assiste actuellement à l'apparition de véhicules propulsés exclusivement grâce à eux, tels que le "tramway sans rail" Blue Tram de Bolloré, qui se recharge en quelques minutes en station avant de parcourir une distance de l'ordre du kilomètre, ou encore un bateau électrique faisant de nombreux allers-retours sur de courtes distances (Ar Vag Tredan). Cependant, le champ d'applications reste limité et il est nécessaire d'augmenter la densité d'énergie des supercondensateurs, c'est-à-dire la quantité d'énergie que l'on peut stocker par unité de masse.

Les supercondensateurs fonctionnent grâce à deux électrodes en carbone plongées dans une solution ionique ou un liquide ionique pur. Il est depuis longtemps établi que c'est l'adsorption d'ions à la surface des électrodes qui permet de stocker l'électricité.

Toutefois, le mécanisme microscopique à l'origine des performances exceptionnelles des carbones dérivés de carbures (CDC) pour le stockage de la charge [3] restait à préciser. Outre la difficulté à décrire correctement une interface entre un liquide ionique et une surface métallique [4], il faut également prendre en compte l'effet du confinement extrême dans ces matériaux. Par simulation moléculaire d'électrodes de structure réaliste et maintenues à potentiel constant, nous étudions les effets du confinement et de la solvatation sur le mécanisme de charge. Nous précisons également la dynamique du processus de charge et faisons le lien avec les modèles utilisés par les électrochimistes.

## 2) Méthodologie

Les simulations moléculaires reposent sur une description d'un système à l'échelle atomique, et permettent d'étudier les propriétés structurales, thermodynamiques et dynamiques de façon complémentaire aux expériences [5]. Une simulation moléculaire est une expérience numérique, au cours de laquelle on prépare le système dans un état initial, puis on observe son évolution, que l'on prédit en résolvant numériquement l'équation de Newton (ou principe fondamental de la dynamique). L'analyse de la trajectoire de tous les atomes permet ensuite de calculer les propriétés, qui sont comparées aux expériences (par exemple la capacité électrique d'un supercondensateur) ou fournissent des informations difficiles à obtenir expérimentalement (par exemple le nombre de coordination d'un ion dans une électrode).

La pertinence des simulations moléculaires dépend dès lors de la façon dont on décrit le système et les interactions entre ses constituants. Au cours de nos travaux, nous prenons en compte deux aspects essentiels souvent négligés : d'une part, une structure réaliste des matériaux d'électrode, qui présente une porosité complexe à l'échelle nanométrique, et d'autre part, la polarisation de l'électrode (dont le potentiel est maintenu constant) par l'électrolyte à sa surface. La structure de l'électrolyte évolue en permanence sous l'effet des fluctuations thermiques, ce qui induit des fluctuations de la distribution de la charge dans l'électrode. Le coût de calcul associé à cette description réaliste de l'électrode est compensé par l'utilisation d'un modèle « à gros-grain » de l'électrolyte (l'hexafluorophosphate de butyl-methyl-imidazolium, ou BMI-PF$_6$), soit sous forme de liquide ionique pur (sans solvant), soit dans un solvant organique, l'acétonitrile.

La comparaison entre des électrodes de graphite et en carbone nanoporeux nous permet de préciser le rôle du confinement, tandis que celle entre le liquide ionique pur et les mêmes ions dans un solvant permet de déterminer celui de la solvatation (==Figure 2==). Dans chaque cas, on évalue d'abord la capacité C du système, définie par le rapport entre la charge Q portée par les électrodes et la différence de potentiel Δψ appliquée aux bornes de la cellule électrochimique modèle. Cette quantité peut être comparée aux expériences. On peut ensuite étudier les propriétés de l'interface électrode-électrolyte à l'échelle moléculaire. Les simulations donnent par exemple accès aux profils de densité cationique et anionique, à l'orientation des molécules, aux nombres de coordination par les autres ions ou par le solvant, etc. Enfin, les propriétés dynamiques peuvent être déterminées, soit par des simulations d'équilibre (coefficient de diffusion, temps de résidence à la surface, etc) ou hors équilibre, au cours de laquelle la réponse à un changement de différence de potentiel est analysée (résistance de l'électrode, etc). Il est ainsi possible de faire le lien avec les modèles de circuits équivalents (ligne à transmission) tels que ceux utilisés par les expérimentateurs.

## 3) Résultats et discussion

Grâce à nos modèles réalistes de supercondensateurs, nous obtenons des résultats pour la capacité des différents systèmes en accord avec les résultats expérimentaux (voir Figure 2). En particulier, l'utilisation d'un solvant permet d'obtenir des capacités comparables au liquide ionique pur tout en travaillant à température ambiante (alors que ce liquide ionique pur doit être utilisé vers 100°C, à cause de sa trop grande viscosité à température ambiante). De plus, les électrodes de CDC conduisent à une augmentation significative de la capacité spécifique (c'est-à-dire par unité de masse). Cette augmentation n'est pas seulement due à la plus grande surface spécifique, et nous avons pu montrer le rôle joué par la microstructure, au-delà de la seule taille de pore moyenne.

L'électrolyte mouille les électrodes même à potentiel nul, ce qui invalide l'idée première selon laquelle c'est l'application d'une différence de potentiel qui induit l'entrée des ions dans l'électrode. De plus, le processus de charge implique l'échange d'ions sans changement notable du volume occupé par le liquide dans l'électrode. Ainsi, des cations sont remplacés par des anions dans l'électrode positive, et réciproquement dans l'électrode négative [6]. L'excès de charge ionique du liquide dans les pores et l'excès de charge électronique de la surface se compensent : on parle d'état super-ionique. Cet échange, observé depuis par des expériences de Résonance Magnétique Nucléaire

(RMN) et de microbalance électrochimique à quartz (EQCM) [7], est accompagné d'une diminution du nombre de coordination des ions, rendue possible par la compensation de la charge des ions par l'électrode. L'efficacité du processus de charge par rapport au cas des électrodes planes de graphite vient du confinement, qui empêche l'apparition du phénomène de « sur-écrantage » qui est défavorable à la séparation des ions à la surface.

Au sein de l'électrode, nous avons pu analyser finement les différents environnements explorés par les ions [8]. Nous avons ainsi proposé quatre types de sites pour l'adsorption des ions, appelés bords, plans, creux et poches, pour des nombres de coordination par les atomes des carbones de l'électrode de plus en plus élevés (voir Figure 3). Nous avons montré que la désolvatation et la charge stockée dans l'électrode par ion sont d'autant plus importantes que le degré de confinement des ions est élevé.

On pourrait s'attendre à ce que les électrodes nanoporeuses de carbone, qui permettent d'obtenir de plus grandes capacités spécifiques que les géométries plus simples, ne sont pas favorables à l'obtention de grandes puissances spécifiques, à cause du temps nécessaire aux ions pour diffuser au sein de l'électrode. Pourtant, les expériences n'indiquent pas une telle tendance. Comment expliquer cette observation à partir du mécanisme à l'échelle moléculaire? Pour y parvenir, nous avons effectué des simulations de dynamique moléculaire au cours desquelles, partant d'un état d'équilibre non chargé, le système est soudainement perturbé par l'application d'une différence de potentiel. On peut ainsi suivre la charge des électrodes, mais également sa répartition dans le matériau, et la corréler à celle des ions dans les pores.

Nous avons ainsi observé que la charge des électrodes de CDC est rapide et hétérogène à l'échelle nanométrique (celle des pores). Les électrodes se chargent progressivement depuis l'interface avec le bulk à mesure que les ions sont échangés entre l'électrode et l'électrolyte. En analysant nos résultats de simulation avec un modèle de circuit équivalent, nous avons extrait la résistance de l'électrode par unité de longueur. Cette dernière intervient, avec la capacité par unité de longueur déterminée par nos simulations d'équilibre, dans le modèle de ligne à transmission utilisé par les électrochimistes pour interpréter leurs données expérimentales. Nous extrapolons ainsi les résultats de simulations à l'échelle de quelques nanomètres et quelques nanosecondes, à des épaisseurs de grains comparables aux matériaux réels. Les temps de charge/décharge obtenus, de l'ordre de la seconde, sont en bon accord avec les résultats expérimentaux.

Le transport des ions à travers les électrodes poreuses est associé à différents processus, couvrant plusieurs échelles de temps [9]: dé/resolvatation (quelques picosecondes), adsorption/désorption à la surface (dizaines à centaines de picosecondes), diffusion à l'intérieur d'un pore et d'un pore à l'autre (nanosecondes). Ces différentes échelles de temps, illustrées sur la Figure 3, dépendent de la différence de potentiel entre les électrodes, qui joue à la fois sur les interactions entre les ions et les surfaces et sur la composition du liquide dans les pores, comme expliqué plus haut.

Enfin, nous avons récemment exploré de nouvelles pistes pour le calcul de la capacité différentielle (dérivée de la charge de l'électrode par rapport à la différence de potentiel) par simulation moléculaire et pour la prédiction des propriétés interfaciales en fonction de la différence de potentiel entre les électrodes. Cette nouvelle stratégie exploite les fluctuations d'équilibre de la charge des électrodes sous l'effet de l'agitation thermique dans l'électrolyte. Nous avons ainsi pu faire le lien entre des pics de la capacité différentielle et des transitions, induites par la différence de potentiel, au sein de l'électrolyte adsorbé [10].

## 4) Conclusion et perspectives

Les simulations moléculaires classiques sont un outil très performant pour l'étude des mécanismes microscopiques à l'origine des excellentes performances des électrodes nanoporeuses de carbone pour le stockage de l'électricité dans les supercondensateurs. Les connaissances acquises à l'échelle moléculaire vont maintenant permettre d'optimiser le choix de la combinaison de structure d'électrode et de composition de l'électrolyte, et ouvrent des perspectives pour le design de nouveaux supercondensateurs encore plus performants.

## Remerciements

Les auteurs remercient leurs collègues Paul Madden à Oxford et Patrice Simon, Pierre-Louis Taberna et Barbara Daffos à Toulouse. La thèse de Céline Merlet a été financée par l'ANR (ANR-2010-BLAN-0933-02) et celle de Clarisse Péan par l'ERC (Patrice Simon, ERC grant agreement 102539). Les résultats ont été obtenus grâce aux ressources de calcul du GENCI (projets c2013096728, x2012096728, x2014096728 et x2015096728), de PRACE (supercalculateur CURIE) et de l'EPSRC (supercalculateur HECTOR).

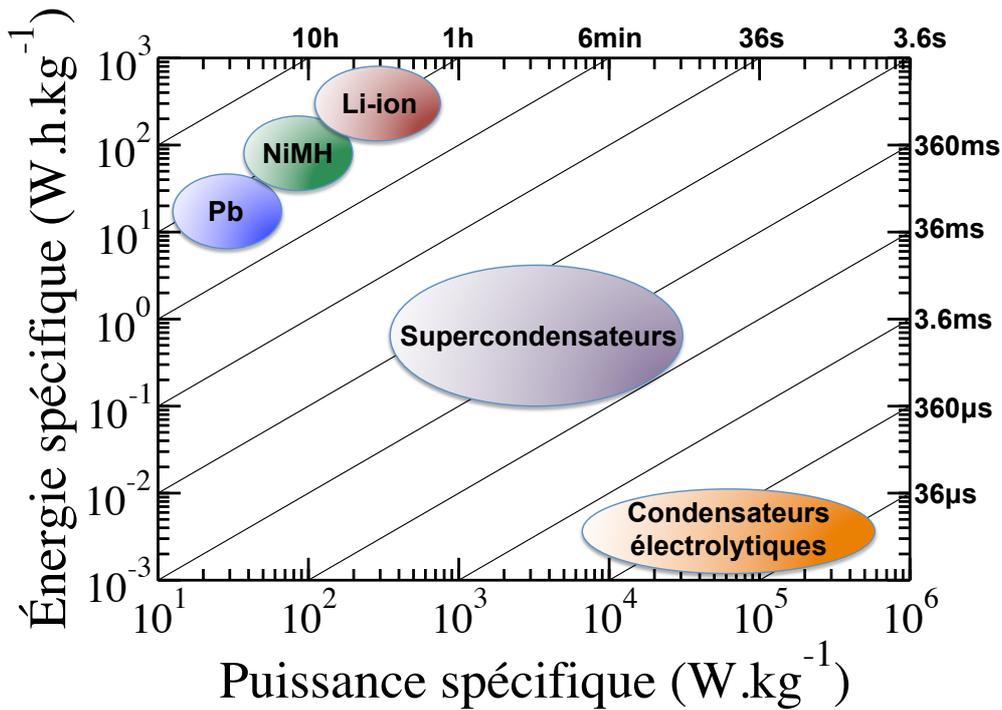

**Fig. 1.** Le diagramme de Ragone représente les différents dispositifs de stockage de l'électricité en fonction de leur puissance spécifique (puissance par unité de masse) et de leur énergie spécifique (énergie par unité de masse). Les supercondensateurs sont ainsi complémentaires des batteries (en haut à gauche) et des condensateurs conventionnels (en bas à droite). La double échelle logarithmique permet de comparer des dispositifs aux performances couvrant plusieurs ordres de grandeur. Les lignes diagonales indiquent le temps caractéristique de charge/décharge.

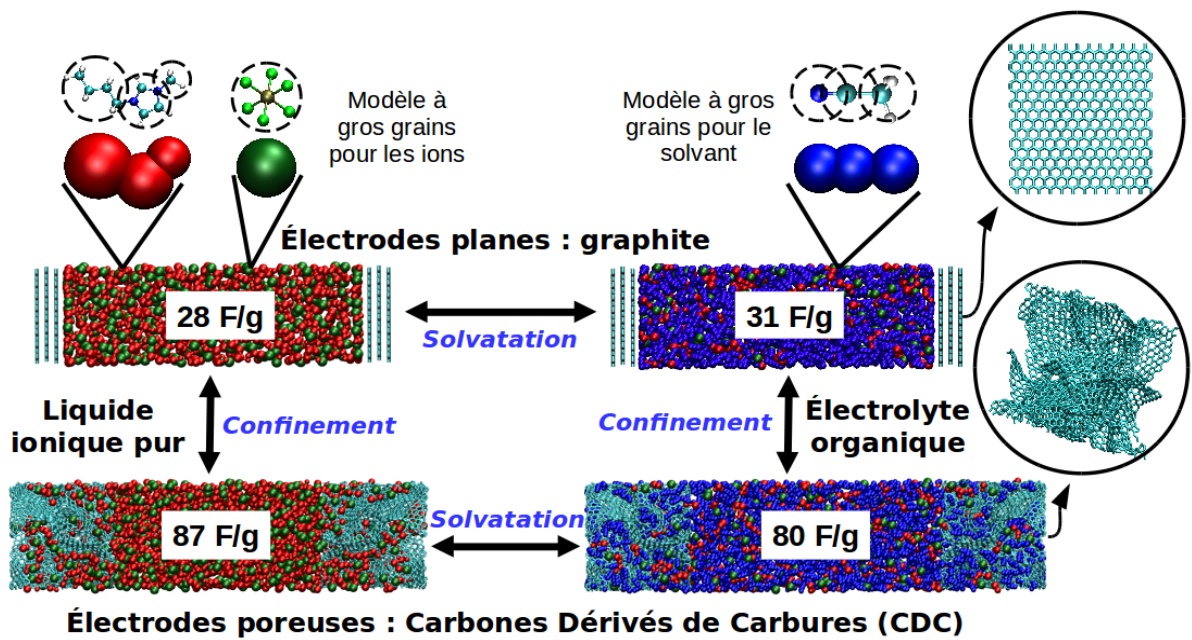

**Fig. 2.** Stratégie de simulation moléculaire (voir texte). Les modèles prennent en compte la structure complexe des électrodes en Carbones Dérivés de Carbures (CDC) ainsi que leur polarisation par l'électrolyte lorsqu'elles sont maintenues à un potentiel constant. Pour les ions et les molécules de solvant de l'électrolyte, des modèles à gros grains sont utilisés.

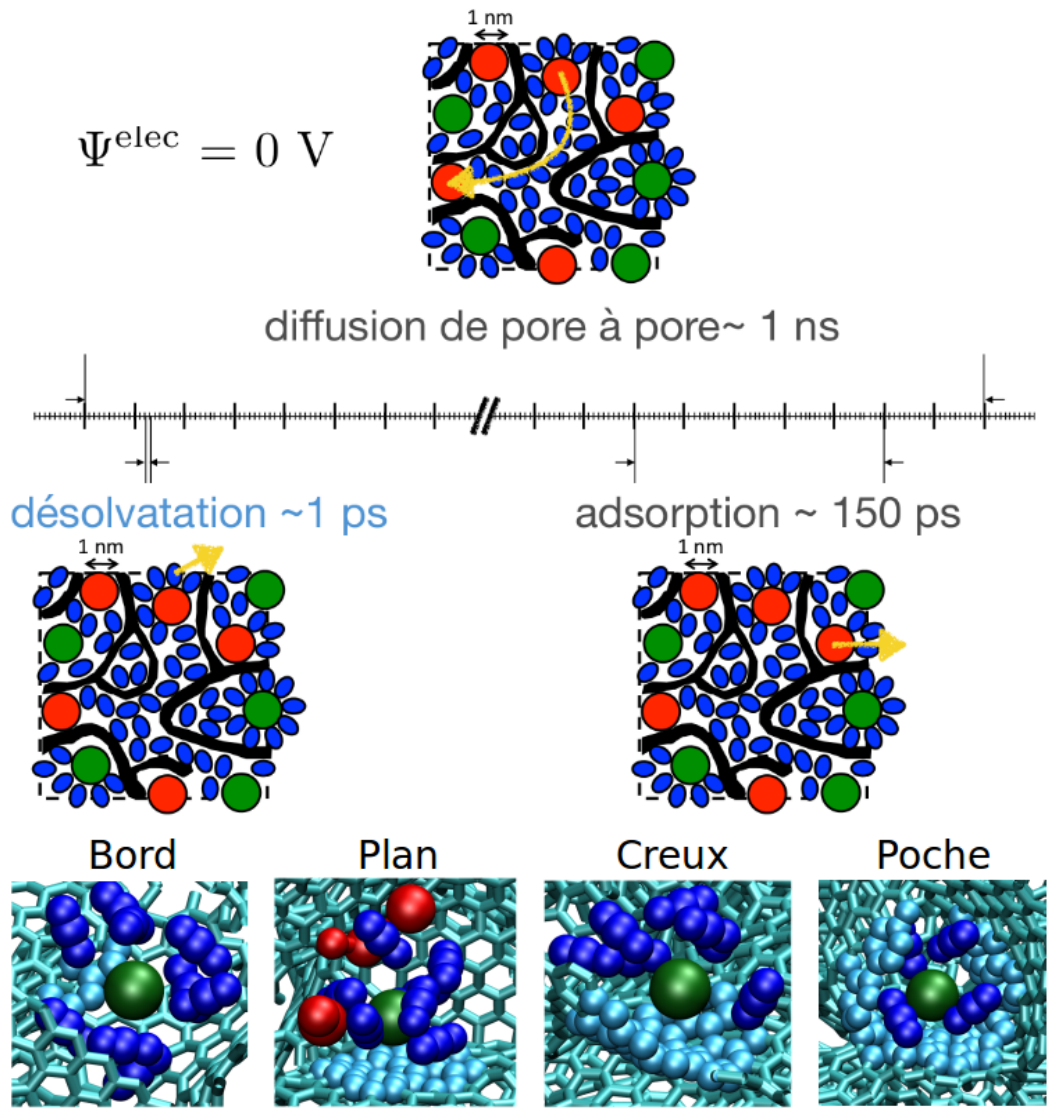

**Fig. 3.** Le transport des ions à travers les électrodes poreuses est associé à différents processus, couvrant plusieurs échelles de temps: dé/resolvatation, adsorption/désorption à la surface, diffusion à l'intérieur d'un pore et d'un pore à l'autre (d'après [9]). La cinétique de ces processus dépend de la différence de potentiel entre les électrodes ainsi que de l'électrode considérée. L'adsorption à la surface conduit ainsi à différents environnements typiques, en fonction du nombre d'atomes de carbone (cyan) en contact avec l'ion (cation en rouge ou anion en vert), et à une variation du nombre de molécules de solvant (en bleu foncé) dans la sphère de solvatation (d'après [8]).